\documentclass[aps,
notitlepage,twocolumn,showpacs,superscriptaddress,nofootinbib,preprintnumbers]{revtex4-2}  
\usepackage{graphicx}  
\usepackage{dcolumn}   
\usepackage{bm}        
\usepackage{amssymb}   
\usepackage{amsmath}
\usepackage[normalem]{ulem}
\usepackage{slashed}
\usepackage{array}
\usepackage{mathtools}
\usepackage[colorlinks=true,citecolor=teal,urlcolor=teal,linkcolor=purple]{hyperref}
\usepackage{cleveref}
\usepackage{tikz}
\usepackage{orcidlink}
\usetikzlibrary{patterns}

\usepackage{soul}

\usepackage{tikz}
\usetikzlibrary{decorations.pathmorphing}
\usetikzlibrary{decorations.markings}
\usetikzlibrary{positioning, shapes, decorations, arrows}
\usetikzlibrary{patterns}

\tikzset{
    wl/.style={line width=1pt},
    graviton/.style={line width=.8pt, -latex,decorate, decoration={snake, segment length=4pt,amplitude=1.8pt, pre length=.15cm, post length=.25cm}},
    gravitonPlain/.style={line width=.8pt,decorate, decoration={snake, segment length=4pt,amplitude=1.8pt, pre length=.05cm}},
    worldlineStatic/.style={black!25, dashed, line width=1pt},
	worldline/.style={black!20, line width=1pt},
	worldlineBold/.style={black, line width=.6pt},
	zUndirected/.style={line width=1pt},
	zParticle/.style={line width=1pt,postaction={decorate},decoration={markings,mark=at position .6 with {\arrow[#1]{latex}}}},
    zParticlePlain/.style={black, line width=1pt},
	zParticle2/.style={line width=1pt,postaction={decorate},decoration={markings,mark=at position .7 with {\arrow[#1]{latex}}}},
	worldlineCut/.style={dotted,line width=1pt,postaction={decorate},decoration={markings,mark=at position .7 with {\arrow[#1]{latex}}}},
	worldlineCut2/.style={dotted,line width=1pt,postaction={decorate},decoration={markings,mark=at position .6 with {\arrow[#1]{latex}}}},
	zParticleF/.style={line width=1pt,postaction={decorate}},
	cscalar/.style={line width=1pt,postaction={decorate},decoration={markings,mark=at position .6 with {\arrow[#1]{latex}}}},
	cscalar2/.style={dotted,line width=.7pt},
	photon/.style={line width =.8pt, decorate, decoration={snake, segment length=4pt, amplitude=1.8pt,  pre length=.1cm, post length=.1cm}},
	photonRed/.style={red, line width =.8pt, decorate, decoration={snake, segment length=4pt, amplitude=1.8pt,  pre length=.1cm, post length=.1cm}},
	cross/.style={cross out, line width =.8pt, draw=black, minimum size=2*(#1-\pgflinewidth), inner sep=0pt, outer sep=0pt},
cross/.default={4pt}
}

\allowdisplaybreaks

\newcommand{\cO}{\mathcal{O}}

\newcommand{\bc}{\bar c}

\def\dd{\delta\!\!\!{}^-\!}

\begin{document}

\title{Classical gravitational scattering with a massive scalar mediator}

\author{Birgitta Biendarra\,\orcidlink{0000-0001-8510-173X}}
\email[]{birgitta.biendarra@hu-berlin.de}
\affiliation{Institut f\"ur Physik, Humboldt-Universit\"at zu Berlin,
 10099 Berlin, Germany}

\author{Kays Haddad\, \orcidlink{0000-0002-1182-2750}}
\email[]{kays.haddad@physik.hu-berlin.de}
\affiliation{Institut f\"ur Physik, Humboldt-Universit\"at zu Berlin,
  10099 Berlin, Germany}

\author{Jan Plefka\, \orcidlink{0000-0003-2883-7825}}
\email[]{jan.plefka@hu-berlin.de}
\affiliation{Institut f\"ur Physik, Humboldt-Universit\"at zu Berlin,
   10099 Berlin, Germany}


\begin{abstract}
We consider the classical scattering of two gravitating compact objects in the presence of a massive scalar mediator, providing a simple model of exotic phenomena.
Through dimensional analysis, we argue that such a process can only be classical in the presence of gravity, a consequence of which is that perturbing in the coupling of the scalar to a worldline is not separate from the post-Minkowskian expansion.
When computing asymptotic observables, the massive mediator complicates the Fourier transforms to impact-parameter space at next-to-leading order.
We reduce these to univariate parametric integrals -- amenable to numerical integration -- and produce analytic results for the linear impulse and the scattering angle to the second post-Minkowskian order.
The scattering angle exhibits a {maximum} when the range of the scalar-mediated force is comparable to the impact parameter, {a possible signature of a massive scalar mediator}.
In the opposite, large-mass regime we uncover a screening effect: {integrating out the scalar sourced by each compact object reduces its gravitational mass} by an amount linear in the scalar's mass.
Both of these phenomena are next-to-leading-order effects.
\end{abstract}

\pacs{}

\preprint{HU-EP-26/26}

\maketitle


\section{Introduction}


The direct detection of gravitational waves (GWs) from a binary-black-hole merger represented a resounding success for the theory of general relativity (GR) \cite{LIGOScientific:2016aoc}.
At the same time, the possibility of observing compact mergers through the lens of GWs opens a new channel for studying open problems in gravitation, cosmology, and astro(particle-)physics.

Several of these problems may be approached by the introduction of a new scalar degree of freedom.
For example, quintessence and k-essence replace the cosmological constant with a scalar field (potentially with non-canonical kinetic term) as the cause of the accelerating expansion of the universe \cite{Peebles:1987ek,Chiba:1999ka}.
Still other models use a scalar field to modify gravity itself, either as a description of dark energy \cite{Brans:1961sx,Tsujikawa2011}, or to explain the dimensionality of Newton's constant \cite{Fujii:1982ms}.

On the astrophysical side, boson stars are stable solutions to the Klein-Gordon-Einstein equations \cite{Das:1963,Kaup:1968zz,Lee:1986ts}, whose mergers may in fact be detectable in current GW observatories \cite{Evstafyeva:2024qvp}.
Various models of these exotic celestial bodies exist, differing by the properties of the scalar field and the potential describing its interactions \cite{Liebling:2012fv}.
While these objects may provide an explanation for dark matter, it is also easy to conceive of the scalar field as describing a stand-alone dark matter particle.
Also, hypothetical ultra-light scalar fields may produce boson clouds around black holes, yielding possibly observable superradiant effects in black-hole inspirals \cite{Baumann:2018vus}.

An action-based approach is convenient in all contexts, in which the Einstein-Hilbert action is extended to include a scalar particle. Modelling the compact object -- be it a
black hole, neutron star, boson star or other exotic -- via a massive worldline effective
theory, one may also introduce direct couplings between the worldline and the scalar.
In this paper, we are concerned with the influence of a massive scalar on observables pertinent to GW observations, either due to binary coalescence involving a boson star or the presence of scalar dark matter.

The classical scattering of two well-separated compact bodies is efficiently captured by an effective worldline theory, in which each body is described by a massive point particle coupled to the gravitational field, with its finite-size structure encoded in higher-dimensional worldline operators \cite{Goldberger:2004jt,Kalin:2020mvi,Mogull:2020sak}.
In the post-Minkowskian (PM) expansion in Newton's constant $G$, the key asymptotic observables -- the impulse, scattering angle, and emitted waveform -- may then be obtained using modern tools from perturbative quantum field theory.
The worldline quantum field theory (WQFT) formalism \cite{Mogull:2020sak,Jakobsen:2022psy,Jakobsen:2023oow,Haddad:2024ebn,Gonzo:2026yha} has proven especially powerful in this regard, systematically incorporating spin, tidal, and radiation reaction effects \cite{Jakobsen:2021smu,Jakobsen:2022fcj,Jakobsen:2022psy,Jakobsen:2022zsx,Haddad:2024ebn,Haddad:2025cmw}.
Building on advanced multi-loop integration techniques, this programme has recently pushed
our resolution of the gravitational two-body problem to the fourth- and fifth-PM (i.e.~$\cO(G^{4})$ and $\cO(G^{5})$)  orders \cite{Dlapa:2022lmu,Dlapa:2023hsl,Driesse:2024xad,Driesse:2024feo,Driesse:2026qiz,Bohnenblust:2026ujk}, see also \cite{Bern:2021dqo, Bern:2021yeh, Damgaard:2023ttc,Bern:2025zno,Bern:2025wyd} for the related amplitude-based approach.

Working within this effective worldline framework, both the bulk and worldline actions may be extended to incorporate the scalar field.
While similar setups have recently been studied in refs.~\cite{Bhattacharyya:2024aeq,Damour:2025oys}, we focus here on establishing analytic control of the perturbation theory at next-to-leading order (NLO).
As such, we consider only the minimal Lagrangian describing a real scalar which interacts gravitationally, and the lowest-dimension operators on the worldline which are relevant for the precision we are after.

Our task requires that we properly organise the perturbative expansion of the new effects, as well as handle complications to integration brought about by the mass of the new mediator.
One innovation in this regard is the realisation that scalar couplings are commensurate with gravitational ones in a post-Minkowskian (PM) expansion.
Consequently, we contend that the purportedly 2PM impulse of ref.~\cite{Bhattacharyya:2024aeq} in fact mixes 2 and 3PM contributions.
A second is that we present compact forms of the Fourier transforms at NLO with the full mass dependence of the mediating scalar.
Univariate parametric integrals are left over, but these are easily handled numerically.
Interestingly, we observe that the case of a massive mediator is not constructible by adding small mass corrections to the massless case, but the latter is accessible as a limit of the massive case. 
Looking instead at the large-mass limit, we observe a screening effect induced by a
scalar-field ``cloud'' surrounding the compact objects, which reduces their effective gravitational mass.

Our final results include the linear impulse and the scattering angle up to 2PM precision.
We evaluate the latter numerically and compare its significance relative to the purely gravitational scattering scenario.
Our computation reveals that scalar-mediated scattering distinguishes itself from purely-gravitational interactions in the form of a resonance in the scattering angle.

We set up our extended WQFT action in \Cref{sec:Action}.
There we also analyse the scales of the problem and argue for the equal relevance of a scalar mediator and a graviton at a fixed PM order.
We present the results of our computation of the linear impulse and the scattering angle up to 2PM in \Cref{sec:Results}, elaborating on our computation of the Fourier transforms with full mass dependence.
We enhance our understanding of scalar-mediated scattering by studying limits of the impulse in \Cref{sec:Limits}.
Plots of the scattering angle in \Cref{sec:ScatteringAngle} quantify the size of the contributions from the massive scalar, as a function both of the scalar's own mass and the impact parameter.
We conclude in \Cref{sec:Conclusions}.

\section{Worldline \& bulk dynamics}\label{sec:Action}

We consider a bulk theory that augments the Einstein-Hilbert action with a minimally coupled
real scalar field,
\begin{align}\label{eq:BulkAction}
    S_{\rm bulk}=\int{\rm d}^{d}x\,\sqrt{-g}\left[\frac{2R}{\kappa^{2}}+\frac{g^{\mu\nu}}{2}\partial_{\mu}\phi\,\partial_{\nu}\phi-\frac{\mu^{2}}{2}\phi^{2}\right],
\end{align}
in a mostly-minus metric convention.
For the purposes of dimensional regularisation, we treat bulk dynamics in $d=4-2\varepsilon$ dimensions.
In this action we use $g$ to denote the metric determinant, $R$ is the Ricci scalar, and $\kappa=\sqrt{32\pi G}$.
We allow for a scalar mass term, with $\mu$, carrying the dimension of inverse length, being the scale of the scalar force's range.
A gauge fixing term for the graviton is suppressed; we work in de Donder gauge.

The graviton and scalar mediators influence the dynamics of a massive body charged under the scalar field, whose effective worldline action (in proper time gauge)
we take to be of the form
\begin{align}\label{eq:WQFTAction2}
    S_{i}=\int{\rm d}\tau_{i}\left[-\frac{m_{i}}{2}g_{\mu\nu}(x_{i})\dot{x}_{i}^{\mu}\dot{x}_{i}^{\nu}+\sum_{j=1}^{2}c_{i,j}\phi(x_{i})^{j}\right],
\end{align}
with $i=1,2$ labelling the two worldlines.

In order to correctly organise the perturbation theory for the system described by \cref{eq:BulkAction,eq:WQFTAction2}, we must understand the scalings of the new coupling constants $c_{i,j}$ along with the mass term $\mu$.
We work with $c=1$, equating time and length, but refrain from
setting  $\hbar=1$ as we are interested in classical physics, so mass $[M]$ and length $[L]$ remain distinct.
It is natural to assign $\tau_{i}$ the dimension of length, leading to the following dimensionalities of fields and couplings (in 4D):
\begin{equation}\label{eq:scalings}
\begin{aligned}
[\phi] = \sqrt{\frac{[M]}{[L]}}\, ,& \quad
[G]= \frac{[L]}{[M]}\, , \\
[\mu] = [L]^{-1}\, , \quad
[c_{i,1}] = &\sqrt{[L][M]}\, \quad
[c_{i,2}] = [L]\, .
\end{aligned}
\end{equation}
As $[\hbar]=[M][L]$, the dimensions of the scalar couplings could be
realised by the scalings $[\mu]\sim m_{\phi}/\hbar$ and $c_{i,j}\sim m_*^{1-j}\hbar^{j/2}$, where $m_*$ is either the mass of worldline $i$ or of the massive scalar, but such choices are poorly motivated in a classical scenario.
{We could in principle introduce a new length scale for the $c_{i,j}$ couplings.
However, naturalness in the presence of gravity calls for the classical scaling}
\begin{align}\label{eq:CoefficientScaling2}
    c_{i,j}= {\bar c}_{i,j} m_i G^{j/2}\, , \quad {\bar c}_{i,j}\in\mathbb{R}\, ,
\end{align}
thereby assuming the coupling of the scalar field to the worldline to be independent of
$\mu$ {(or any other length scale)}.\footnote{Note that there is no analogous classical scaling for $\mu$, as $\mu\sim m_{\phi} G^{k}$ is not possible for any $k$. Hence, $\mu$ must set a new inverse length scale in our theory.}

The linear coupling $\bar{c}_{i,1}$ may be thought of as the scalar charge of the
$i^{\rm th}$ massive body.
As we will see in \cref{eq:LOImpulseEvaluated}, taking the charges of the two bodies to be of the same (opposite) sign produces an attractive (repulsive) scalar force at the leading order, while NLO corrections yield a mixed profile of attractive versus repulsive contributions depending on the signs of the $\bar{c}_{i,j}$.

\Cref{eq:CoefficientScaling2} has an unexpected implication: replacing a graviton leg in a vertex with a scalar one does not change that vertex's PM order, i.e.~power of $G$.
As such, our goal of computing $\phi$-mediated scattering up to NLO requires us to compute diagrams with one scalar, one scalar and one graviton, and two scalars exchanged.
Thus, truncating the operators in \cref{eq:WQFTAction2} at second order in the mediating scalar accounts for all contributions relevant to our targeted precision.

With these considerations, we may now employ WQFT perturbation theory as developed in ref.~\cite{Mogull:2020sak}.
Specifically, the metric is perturbed around Minkowski space, $g_{\mu\nu}=\eta_{\mu\nu}+\kappa\,h_{\mu\nu}$, and the worldlines around straight trajectories, $x_{i}^{\mu}(\tau)=b_{i}^{\mu}+v_{i}^\mu \tau_{i}+z_{i}^{\mu}(\tau_{i})$.
We take the background configuration of the scalar mediator to be vanishing.
Classical dynamics are extracted from diagrams with no closed loops: each pseudoloop at subleading PM orders must be enclosed by exactly one static worldline.

The WQFT Feynman rules follow straightforwardly. Next to the known purely gravitational
worldline and bulk rules (see refs.~\cite{Mogull:2020sak,Driesse:2024feo}), the scalar adds
the momentum-space propagator
\begin{equation}
     \langle \phi(q) \phi(-q)\rangle =\,
        \begin{tikzpicture}[line cap=round,line join=round,baseline=-0.1cm]
             \node (A3) at (0.5,-.25) {$q$};
               \path [draw=black, cscalar2] (0,0) -- (1,0);
    \end{tikzpicture}\, =
    \frac{i}{q^{2}-\mu^{2}}
    ,
\end{equation}
denoting the massive scalar with a dotted line.
New worldline-scalar interaction vertices emerge from \cref{eq:WQFTAction2}:
\begin{align}
\begin{tikzpicture}[line cap=round,line join=round,baseline=-0.4cm]
    \path [draw=black!40, worldlineStatic] (0,0) -- (1.2,0);
     \path [draw=black, cscalar2] (0.6,0) -- (0.6, -0.8);
    \fill [draw=black] (0.6,0) circle (1.75pt);
\end{tikzpicture}
\quad
\begin{tikzpicture}[line cap=round,line join=round,baseline=-0.4cm]
    \path [draw=black!40, worldlineStatic] (0,0) -- (1.2,0);
     \path [draw=black, zParticle2] (0.6,0) -- (1.2,0);
     \path [draw=black, cscalar2] (0.6,0) -- (0.6, -0.8);
    \fill [draw=black] (0.6,0) circle (1.75pt);
\end{tikzpicture}
\quad
\begin{tikzpicture}[line cap=round,line join=round,baseline=-0.4cm]
    \path [draw=black, zParticle2] (0,0) -- (0.6,0);
     \path [draw=black, zParticle2] (0.6,0) -- (1.2,0);
     \path [draw=black, cscalar2] (0.6,0) -- (0.6, -0.8);
    \fill [draw=black] (0.6,0) circle (1.75pt);
\end{tikzpicture}
&\sim \bc_{i,1} m_{i} G^{1/2}
\\
\begin{tikzpicture}[line cap=round,line join=round,baseline=-0.4cm]
    \path [draw=black!40, worldlineStatic] (0,0) -- (1.2,0);
     \path [draw=black, cscalar2] (0.6,0) -- (0.2, -0.8);
       \path [draw=black, cscalar2] (0.6,0) -- (1.0, -0.8);
    \fill [draw=black] (0.6,0) circle (1.75pt);
\end{tikzpicture}
\quad
\begin{tikzpicture}[line cap=round,line join=round,baseline=-0.4cm]
    \path [draw=black!40, worldlineStatic] (0,0) -- (1.2,0);
     \path [draw=black, zParticle2] (0.6,0) -- (1.2,0);
     \path [draw=black, cscalar2] (0.6,0) -- (0.2, -0.8);
       \path [draw=black, cscalar2] (0.6,0) -- (1.0, -0.8);
    \fill [draw=black] (0.6,0) circle (1.75pt);
\end{tikzpicture}
\sim & \,\,\bc_{i,2} m_{i} G.
\end{align}
In these and other diagrams in this paper, static (i.e.~the particle $i$ background,
$b_{i}^{\mu}+v_{i}^{\mu}\tau$) and fluctuating worldlines (i.e.~the worldline deflection mode $z_{i}^{\mu}$) are represented by grey dashed and black solid lines respectively.
Arrows on the latter indicate the flow of causality, necessarily oriented towards the outgoing state \cite{Jakobsen:2022psy}.
At 2PM order we will not need higher $z_{i}$-emissions.
Finally, we encounter a novel bulk scalar-graviton vertex from \cref{eq:BulkAction}:
\begin{equation}
\begin{tikzpicture}[line cap=round,line join=round,baseline=-1.0cm]
        \path [draw=black, photon] (0,-.375) -- (-0.5, -0.925);
        \fill [draw=black] (-0.5,-.925) circle (1.75pt);
        \path [draw=black, cscalar2] (-1,-.375) -- (-0.5, -0.925);
        \path [draw=black, cscalar2] (-0.5,-0.925) -- (-0.5, -1.425);
\end{tikzpicture}
\sim G^{1/2}\, .
\end{equation}
Again, we do not need higher graviton emissions at the NLO considered.
We provide the explicit vertex rules in appendix \ref{app:FeynmanRules}.

A typical description of boson stars uses a solitonic potential that includes quartic and sextic scalar interactions \cite{Liebling:2012fv}.
Our present bulk action omits these effects, but they may enter for the first time at two-loop order, such that they are irrelevant at the precision considered here.
Moreover, we consider the simpler case of a real rather than a complex scalar; the latter involves complex sources which couple the scalar to worldlines, as formulated in ref.~\cite{Damour:2025oys}, which introduce new scales into the problem.
We leave the investigation of this more involved setup to future work.

{Ref.~\cite{Bhattacharyya:2024aeq} considered an identical naturalness argument for the worldline-scalar
couplings \eqref{eq:CoefficientScaling2} by deforming the mass $m_{i}\to m_{i}(\phi)= m_{i} ( 1 + s_{i}\kappa \phi + g_{i}\kappa^2 \phi^{2}+\ldots) $, which for the linear in $\phi$ term
we have confirmed produces the same impulse to NLO precision as \cref{eq:WQFTAction2}.
Yet,  Ref.~\cite{Bhattacharyya:2024aeq}  missed the inclusion of the
\emph{non-fluctuating} worldline-scalar-scalar vertex, cf.~\eqref{eq:A4},  in their analysis, which makes a comparison at 2PM order spurious. Additionally,
the powers of $\kappa$ in $m_i(\phi)$ are omitted in the PM counting of the diagrams
evalauted in \cite{Bhattacharyya:2024aeq}.}

\section{Analytic linear impulse to 2PM}\label{sec:Results}

The linear impulse experienced by a compact object is described in WQFT by the one-point functions involving a deflection mode in the final state,
\begin{align}
    \Delta p_{i}^{\rho}=-m_{i}\,\omega^{2}\langle z_{i}^{\rho}(-\omega)\rangle|_{\omega\rightarrow0},
\end{align}
where
\begin{align}
    \langle z_{i}^\rho(-\omega)\rangle=\vcenter{\hbox{
        \begin{tikzpicture}[line cap=round,line join=round,baseline=-1.8cm]
        \node (A1) at (-2.7,.0) {$z_{i}^\rho(-\omega)$};
        \node (A2) at (-2.7,-.6) {$\omega\rightarrow$};
        \path [draw=black!40, worldlineStatic] (-4.3,-.375) -- (-3.5,-.375);
        \path [draw=black, zParticle2] (-3.5,-.375) -- (-2.7,-.375);
        \path [draw=black!40, worldlineStatic] (-4.3,-1.125) -- (-2.7,-1.125);
        \draw[fill=white] (-3.5,-0.75) ellipse (.24cm and .45cm);
        \draw[pattern=north east lines] (-3.5,-0.75) ellipse (.24cm and .45cm);
    \end{tikzpicture}}},
\end{align}
with the blob depicting all possible diagrams at any loop order.

At leading order, there is only one diagram constituting the impulse from the exchange of a scalar particle,
\begin{align}\label{eq:1PMImpulseDiagram}
    \langle z_{1}^\rho(-\omega)\rangle\big|_{\rm 1PM}^{\rm scalar}=\vcenter{\hbox{
        \begin{tikzpicture}[line cap=round,line join=round,baseline=-1.8cm]
        \node (A1) at (-2.7,.0) {$z_{1}^\rho(-\omega)$};
        \node (A2) at (-2.7,-.6) {$\omega\rightarrow$};
        \node (A3) at (-3.8,-.75) {$q\downarrow$};
        \path [draw=black!40, worldlineStatic] (-4.3,-.375) -- (-3.5,-.375);
        \path [draw=black, zParticle] (-3.5,-.375) -- (-2.7,-.375);
        \path [draw=black, cscalar2] (-3.5,-.375) -- (-3.5, -1.125);
        \path [draw=black!40, worldlineStatic] (-4.3,-1.125) -- (-2.7,-1.125);
        \fill [draw=black] (-3.5,-.375) circle (1.75pt);
        \fill [draw=black] (-3.5, -1.125) circle (1.75pt);
    \end{tikzpicture}}}.
\end{align}
Our convention for the flow of the transfer momentum $q^\mu$ is indicated.
Evaluating this diagram for the impulse felt by worldline 1 gives
\begin{align}\label{eq:LOImpulse}
    \Delta p^{(1),\rho}_{1}=-Gm_{1}m_{2}\bc_{1,1}\bc_{2,1}\frac{\partial}{\partial b_\rho}\mathcal{F}^{(1)}(\mu|b|),
\end{align}
with
\begin{align}\label{eq:TreeLevelFTMomentumSpace}
      \mathcal{F}^{(1)}(\mu|b|)&=\int_{q}\frac{1}{\mu^{2}-q^{2}}.
\end{align}
We have abbreviated the Fourier transform as
\begin{align*}
    \int_{q}=\int\frac{{\rm d}^{d}q}{(2\pi)^{d}}\dd(v_{1}\cdot q)\dd(v_{2}\cdot q)e^{i q\cdot b}.
\end{align*}
Though not apparent before integration, we will see shortly that the transform only depends on the scales $\mu$ and $|b|=\sqrt{-b^{2}}$ through the magnitude of the product $\mu b^{\rho}\equiv\beta^{\rho}$.
The impact parameter,
\begin{align}\label{eq:ImpactParameter}
    b^\rho\equiv {P^{\rho}}_{\nu}\left(b^\nu_1-b^\nu_2\right),
\end{align}
is projected onto the plane orthogonal to $v_{1,2}^{\mu}$ by
\begin{align}\label{eq:Projector}
    P^{\rho\nu}=\eta^{\rho\nu}-v_{1}^{\rho}\Check{v}_{1}^{\nu}-v_{2}^{\rho}\Check{v}_{2}^{\nu}.
\end{align}
This involves the dual velocities
\begin{align}
    \check{v}_{1}^{\rho}=\frac{\gamma v_2^\rho-v_1^\rho}{\gamma^2-1},\quad \check{v}_{2}^{\rho}=\frac{\gamma v_1^\rho-v_2^\rho}{\gamma^2-1},
\end{align}
where $\gamma=v_{1}\cdot v_{2}$.
We may evaluate this Fourier transform without much difficulty using methods from either of refs.~\cite{Kosower:2018adc,Brunello:2025eso}, giving
\begin{align}\label{eq:TreeLevelFT}
       \mathcal{F}^{(1)}(|\beta|)=\frac{K_{0}(|\beta|)}{2\pi\sqrt{\gamma^{2}-1}},
\end{align}
where $K_{\nu}(z)$ are modified Bessel functions of the second kind.
With this, \cref{eq:LOImpulse} becomes
\begin{align}\label{eq:LOImpulseEvaluated}
    \Delta p^{(1),\rho}_{1}=-\frac{Gm_{1}m_{2}}{2\pi|b|}\frac{\bc_{1,1}\bc_{2,1}}{\sqrt{\gamma^{2}-1}}\beta^{\rho}K_{1}(|\beta|),
\end{align}
in agreement with refs.~\cite{Bhattacharyya:2024aeq,Damour:2025oys}.
When the $\bar{c}_{i,1}$ have the same sign this impulse points in the direction of particle 2, indicating that the scalar exchange induces an attractive force.
We remind the reader that, although there are no gravitons in \cref{eq:1PMImpulseDiagram}, \cref{eq:CoefficientScaling2} means this contribution is relevant at 1PM.

\begin{figure*}[t]
\centering


\resizebox{\linewidth}{!}{$
\begin{array}{ccccccc}

\vcenter{\hbox{
\begin{tikzpicture}[line cap=round,line join=round,baseline=-1.8cm]
    \path [draw=black!40, worldlineStatic] (-5,-.375) -- (-3.5,-.375);
    \path [draw=black!40, worldlineStatic] (-5,-1.425) -- (-2.7,-1.425);
    \path [draw=black, zParticle] (-3.5,-.375) -- (-2.7,-.375);
    \fill [draw=black] (-3.5,-.375) circle (1.75pt);
    \path [draw=black, photon] (-3.5,-.375) -- (-3.5, -1.425);
    \fill [draw=black] (-3.5,-1.425) circle (1.75pt);
    \path [draw=black, zParticle2] (-4.5,-.375) -- (-3.5,-.375);
    \fill [draw=black] (-4.5,-.375) circle (1.75pt);
    \path [draw=black, cscalar2] (-4.5,-.375) -- (-4.5, -1.425);
    \fill [draw=black] (-4.5,-1.425) circle (1.75pt);
\end{tikzpicture}}}
&

\vcenter{\hbox{
\begin{tikzpicture}[line cap=round,line join=round,baseline=-1.8cm]
    \path [draw=black!40, worldlineStatic] (-5,-.375) -- (-3.5,-.375);
    \path [draw=black!40, worldlineStatic] (-5,-1.425) -- (-2.7,-1.425);
    \path [draw=black, zParticle] (-3.5,-.375) -- (-2.7,-.375);
    \fill [draw=black] (-3.5,-.375) circle (1.75pt);
    \path [draw=black, cscalar2] (-3.5,-.375) -- (-3.5, -1.425);
    \fill [draw=black] (-3.5,-1.425) circle (1.75pt);
    \path [draw=black, zParticle2] (-4.5,-.375) -- (-3.5,-.375);
    \fill [draw=black] (-4.5,-.375) circle (1.75pt);
    \path [draw=black, photon] (-4.5,-.375) -- (-4.5, -1.425);
    \fill [draw=black] (-4.5,-1.425) circle (1.75pt);
\end{tikzpicture}}}
&

\vcenter{\hbox{
\begin{tikzpicture}[line cap=round,line join=round,baseline=-1.8cm]
    \path [draw=black!40, worldlineStatic] (-5,-.375) -- (-3.5,-.375);
    \path [draw=black!40, worldlineStatic] (-5,-1.425) -- (-2.7,-1.425);
    \path [draw=black, zParticle] (-3.5,-.375) -- (-2.7,-.375);
    \fill [draw=black] (-3.5,-.375) circle (1.75pt);
    \path [draw=black, photon] (-3.5,-.375) -- (-3.5, -1.425);
    \fill [draw=black] (-3.5,-1.425) circle (1.75pt);
    \path [draw=black, zParticle2] (-4.5,-1.425) -- (-3.5,-1.425);
    \fill [draw=black] (-4.5,-.375) circle (1.75pt);
    \path [draw=black, cscalar2] (-4.5,-.375) -- (-4.5, -1.425);
    \fill [draw=black] (-4.5,-1.425) circle (1.75pt);
\end{tikzpicture}}}
&

\vcenter{\hbox{
\begin{tikzpicture}[line cap=round,line join=round,baseline=-1.8cm]
    \path [draw=black!40, worldlineStatic] (-5,-.375) -- (-3.5,-.375);
    \path [draw=black!40, worldlineStatic] (-5,-1.425) -- (-2.7,-1.425);
    \path [draw=black, zParticle] (-3.5,-.375) -- (-2.7,-.375);
    \fill [draw=black] (-3.5,-.375) circle (1.75pt);
    \path [draw=black, cscalar2] (-3.5,-.375) -- (-3.5, -1.425);
    \fill [draw=black] (-3.5,-1.425) circle (1.75pt);
    \path [draw=black, zParticle2] (-4.5,-1.425) -- (-3.5, -1.425);
    \fill [draw=black] (-4.5,-.375) circle (1.75pt);
    \path [draw=black, photon] (-4.5,-.375) -- (-4.5, -1.425);
    \fill [draw=black] (-4.5,-1.425) circle (1.75pt);
\end{tikzpicture}}}
&

\vcenter{\hbox{
\begin{tikzpicture}[line cap=round,line join=round,baseline=-1.8cm]
        \path [draw=black!40, worldlineStatic] (-5,-.375) -- (-3.5,-.375);
        \path [draw=black!40, worldlineStatic] (-5,-1.425) -- (-2.7,-1.425);
        \path [draw=black, zParticle] (-3.5,-.375) -- (-2.7,-.375);
        \fill [draw=black] (-3.5,-.375) circle (1.75pt);
        \path [draw=black, photon] (-3.5,-.375) -- (-4, -0.925);
        \fill [draw=black] (-4.5,-.375) circle (1.75pt);
        \path [draw=black, cscalar2] (-4.5,-.375) -- (-4, -0.925);
        \fill [draw=black] (-4,-0.925) circle (1.75pt);
        \path [draw=black, cscalar2] (-4,-0.925) -- (-4, -1.425);
        \fill [draw=black] (-4,-1.425) circle (1.75pt);
\end{tikzpicture}}}
&

\vcenter{\hbox{
\begin{tikzpicture}[line cap=round,line join=round,baseline=-1.8cm]
        \path [draw=black!40, worldlineStatic] (-5,-.375) -- (-3.5,-.375);
        \path [draw=black!40, worldlineStatic] (-5,-1.425) -- (-2.7,-1.425);
        \path [draw=black, zParticle] (-3.5,-.375) -- (-2.7,-.375);
        \fill [draw=black] (-3.5,-.375) circle (1.75pt);
        \path [draw=black, cscalar2] (-3.5,-.375) -- (-4, -0.925);
        \fill [draw=black] (-4.5,-.375) circle (1.75pt);
        \path [draw=black, photon] (-4.5,-.375) -- (-4, -0.925);
        \fill [draw=black] (-4,-0.925) circle (1.75pt);
        \path [draw=black, cscalar2] (-4,-0.925) -- (-4, -1.425);
        \fill [draw=black] (-4,-1.425) circle (1.75pt);
\end{tikzpicture}}}
&

\vcenter{\hbox{
\begin{tikzpicture}[line cap=round,line join=round,baseline=-1.8cm]
        \path [draw=black!40, worldlineStatic] (-5,-.375) -- (-3.5,-.375);
        \path [draw=black!40, worldlineStatic] (-5,-1.425) -- (-2.7,-1.425);
        \path [draw=black, zParticle] (-4,-.375) -- (-2.7,-.375);
        \fill [draw=black] (-4,-.375) circle (1.75pt);
        \path [draw=black, cscalar2] (-4,-.375) -- (-4, -0.925);
        \fill [draw=black] (-4,-0.925) circle (1.75pt);
        \path [draw=black, cscalar2] (-4,-0.925) -- (-4.5, -1.425);
        \fill [draw=black] (-4.5,-1.425) circle (1.75pt);
        \path [draw=black, photon] (-4,-0.925) -- (-3.5, -1.425);
        \fill [draw=black] (-3.5,-1.425) circle (1.75pt);
\end{tikzpicture}}}

\end{array}
$}

\caption{\label{fig:Family1}The seven diagrams comprising family 1.
All diagrams are proportional to $\kappa^2 c_{1,1}c_{2,1}$, and involve loop integrals with one massless graviton and one massive scalar propagator.}

\end{figure*}
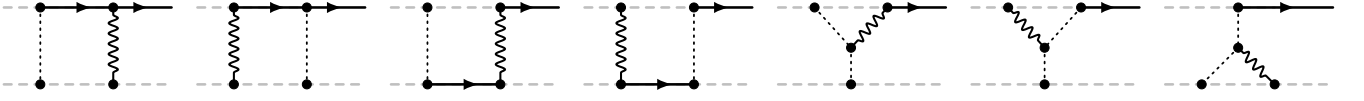

At NLO it is convenient to divide the computation into two families based on the pertinent master integrals.
Let us define the propagators
\begin{subequations}\label{eq:MediatorPropagators}
\begin{align}
    D_{1,\mu}&=\ell^{2}-\mu^{2}+i0^+ \\
    D_{2,\mu}&=(\ell-q)^{2}-\mu^{2}+i0^+.
\end{align}
\end{subequations}
Diagrams in the first family involve integrals with at most one massive propagator, which, in terms of \cref{eq:MediatorPropagators}, are of the form
\begin{align}
    I_{n_{1}n_{2}n_{3}}^{(1,2),\pm}=\tilde{\mu}^{2\varepsilon}\int\frac{{\rm d}^{d}\ell}{(2\pi)^{d}}\frac{\dd(v_{1,2}\cdot \ell)}{(v_{2,1}\cdot\ell\pm i0^+)^{n_{1}}D_{1,0}^{n_{2}}D_{2,\mu}^{n_{3}}},
\end{align}
with the regularisation scale $\tilde{\mu}^{2}=e^{\gamma_{\rm E}}\mu_{\rm DR}^{2}/4\pi$.
In this case the pseudoloop is enclosed by one graviton and one scalar line, as depicted in \cref{fig:Family1}.

The second family accounts for diagrams in which the pseudoloop is enclosed by two scalar lines, falling into the integral family
\begin{align}
    J_{n_{1}n_{2}n_{3}}^{(1,2),\pm}=\tilde{\mu}^{2\varepsilon}\int\frac{{\rm d}^{d}\ell}{(2\pi)^{d}}\frac{\dd(v_{1,2}\cdot \ell)}{(v_{2,1}\cdot\ell\pm i0^+)^{n_{1}}D_{1,\mu}^{n_{2}}D_{2,\mu}^{n_{3}}}.
\end{align}
An additional division of this family is meaningful, with one subfamily, labelled 2a, involving the coupling $\kappa$ and the other, subfamily 2b, involving only the $c_{i,j}$.
Diagrams contributing to family 2 are drawn in \cref{fig:Family2}, with the division into subfamilies also indicated there.
Thus, in all, there are three contributions to the NLO impulse which cannot mix with one another:
\begin{align}
    \Delta p_{1}^{(2),\rho}=\Delta p_{1}^{(2),\rho}\big|_{1}+\Delta p_{1}^{(2),\rho}\big|_{2a}+\Delta p_{1}^{(2),\rho}\big|_{2b},
\end{align}
having denoted contributions from each diagram family.

\begin{figure*}[t]
\centering

\begin{minipage}{0.34\textwidth}
\centering
\resizebox{5cm}{!}{$
\begin{array}{cc cc}

\vcenter{\hbox{
\begin{tikzpicture}[line cap=round,line join=round,baseline=-1.8cm]
        \path [draw=black!40, worldlineStatic] (-5,-.375) -- (-3.5,-.375);
        \path [draw=black!40, worldlineStatic] (-5,-1.425) -- (-2.7,-1.425);
        \path [draw=black, zParticle] (-4,-.375) -- (-2.7,-.375);
        \fill [draw=black] (-4,-.375) circle (1.75pt);
        \path [draw=black, photon] (-4,-.375) -- (-4, -0.925);
        \fill [draw=black] (-4,-0.925) circle (1.75pt);
        \path [draw=black, cscalar2] (-4,-0.925) -- (-4.5, -1.425);
        \fill [draw=black] (-4.5,-1.425) circle (1.75pt);
        \path [draw=black, cscalar2] (-4,-0.925) -- (-3.5, -1.425);
        \fill [draw=black] (-3.5,-1.425) circle (1.75pt);
\end{tikzpicture}}}
&\,

\vcenter{\hbox{
\begin{tikzpicture}[line cap=round,line join=round,baseline=-1.8cm]
        \path [draw=black!40, worldlineStatic] (-5,-.375) -- (-3.5,-.375);
        \path [draw=black!40, worldlineStatic] (-5,-1.425) -- (-2.7,-1.425);
        \path [draw=black, zParticle] (-3.5,-.375) -- (-2.7,-.375);
        \fill [draw=black] (-3.5,-.375) circle (1.75pt);
        \path [draw=black, cscalar2] (-3.5,-.375) -- (-4, -0.925);
        \fill [draw=black] (-4.5,-.375) circle (1.75pt);
        \path [draw=black, cscalar2] (-4.5,-.375) -- (-4, -0.925);
        \fill [draw=black] (-4,-0.925) circle (1.75pt);
        \path [draw=black, photon] (-4,-0.925) -- (-4, -1.425);
        \fill [draw=black] (-4,-1.425) circle (1.75pt);
\end{tikzpicture}}}
&

\end{array}
$}

{\small (a)}
\end{minipage}%
\hfill
\begin{minipage}{0.62\textwidth}
\centering
\resizebox{10cm}{!}{$
\begin{array}{cccc}

\vcenter{\hbox{
\begin{tikzpicture}[line cap=round,line join=round,baseline=-1.8cm]
        \path [draw=black!40, worldlineStatic] (-5,-.375) -- (-3.5,-.375);
        \path [draw=black!40, worldlineStatic] (-5,-1.425) -- (-2.7,-1.425);
        \path [draw=black, zParticle] (-3.5,-.375) -- (-2.7,-.375);
        \fill [draw=black] (-3.5,-.375) circle (1.75pt);
        \path [draw=black, cscalar2] (-3.5,-.375) -- (-4, -1.425);
        \fill [draw=black] (-4.5,-.375) circle (1.75pt);
        \path [draw=black, cscalar2] (-4.5,-.375) -- (-4, -1.425);
        \fill [draw=black] (-4,-1.425) circle (1.75pt);
\end{tikzpicture}}}
&

\vcenter{\hbox{
\begin{tikzpicture}[line cap=round,line join=round,baseline=-1.8cm]
        \path [draw=black!40, worldlineStatic] (-5,-.375) -- (-4,-.375);
        \path [draw=black!40, worldlineStatic] (-5,-1.425) -- (-2.7,-1.425);
        \path [draw=black, zParticle] (-4,-.375) -- (-2.7,-.375);
        \fill [draw=black] (-3.5,-1.425) circle (1.75pt);
        \path [draw=black, cscalar2] (-4,-.375) -- (-3.5, -1.425);
        \fill [draw=black] (-4,-.375) circle (1.75pt);
        \path [draw=black, cscalar2] (-4,-.375) -- (-4.5, -1.425);
        \fill [draw=black] (-4.5,-1.425) circle (1.75pt);
\end{tikzpicture}}}
&

\vcenter{\hbox{
\begin{tikzpicture}[line cap=round,line join=round,baseline=-1.8cm]
        \path [draw=black!40, worldlineStatic] (-5,-.375) -- (-3.5,-.375);
        \path [draw=black!40, worldlineStatic] (-5,-1.425) -- (-2.7,-1.425);
        \path [draw=black, zParticle] (-3.5,-.375) -- (-2.7,-.375);
        \fill [draw=black] (-3.5,-.375) circle (1.75pt);
        \path [draw=black, cscalar2] (-3.5,-.375) -- (-3.5, -1.425);
        \fill [draw=black] (-3.5,-1.425) circle (1.75pt);
        \path [draw=black, zParticle2] (-4.5,-.375) -- (-3.5,-.375);
        \fill [draw=black] (-4.5,-.375) circle (1.75pt);
        \path [draw=black, cscalar2] (-4.5,-.375) -- (-4.5, -1.425);
        \fill [draw=black] (-4.5,-1.425) circle (1.75pt);
\end{tikzpicture}}}
&

\vcenter{\hbox{
\begin{tikzpicture}[line cap=round,line join=round,baseline=-1.8cm]
        \path [draw=black!40, worldlineStatic] (-5,-.375) -- (-3.5,-.375);
        \path [draw=black!40, worldlineStatic] (-5,-1.425) -- (-2.7,-1.425);
        \path [draw=black, zParticle] (-3.5,-.375) -- (-2.7,-.375);
        \fill [draw=black] (-3.5,-.375) circle (1.75pt);
        \path [draw=black, cscalar2] (-3.5,-.375) -- (-3.5, -1.425);
        \fill [draw=black] (-3.5,-1.425) circle (1.75pt);
        \path [draw=black, zParticle2] (-4.5,-1.425) -- (-3.5,-1.425);
        \fill [draw=black] (-4.5,-.375) circle (1.75pt);
        \path [draw=black, cscalar2] (-4.5,-.375) -- (-4.5, -1.425);
        \fill [draw=black] (-4.5,-1.425) circle (1.75pt);
\end{tikzpicture}}}

\end{array}
$}

{\small (b)}
\end{minipage}

\caption{\label{fig:Family2} The two diagrams in family 2a (left) and the four in family 2b (right).
Contributions to the impulse from the former are proportional to $\kappa^2 c_{i,1}^2$, while the latter 
are free of $\kappa$ and at least cubic in the coefficients $c_{i,j}$.
Both subfamilies are united in their dependence on loop integrals with two massive scalar propagators.}

\end{figure*}

Upon assembling the diagrams in \cref{fig:Family1,fig:Family2}, integration-by-parts identities, implemented through \texttt{Kira} \cite{Maierhofer:2017gsa,Klappert:2020nbg,Lange:2025fba}, reduce the loop integrals appearing in either family to a set of master integrals.
In total, four master integrals are needed at one-loop order, all of which are amenable to integration through Feynman or Schwinger parameters.
We list these integrals, evaluated in dimensional regularisation, in \Cref{app:LoopIntegrals}.

\begin{widetext}
    \noindent Performing the loop integrals, the impulse takes the explicit form
    \begin{align}        \Delta p_{1}^{(2),\rho}\big|_{1}&=\frac{G^{2}m_{1}m_{2}}{|b|^{2}}\frac{{\bc}_{1,1}{\bc}_{2,1}(m_1+m_2)}{\gamma^{2}-1}|\beta|\beta^{\rho}\frac{\partial}{\partial|\beta|}\left[\left(2 \gamma ^2-1\right)\mathcal{F}^{(1)}(|\beta|)+(\gamma ^2 -1)\Box_{\beta}\mathcal{F}_{1}^{(2)}(|\beta|)\right]\notag \\
        \Delta p_{1}^{(2),\rho}\big|_{2a}&=\frac{G^{2}m_1 m_{2}}{2|b|^{2}}(\bc_{1,1}^{2}m_{1}+\bc_{2,1}^{2}m_{2})|\beta|\beta^\rho\left[\frac{\sqrt{\gamma^2-1}}{2\pi|\beta|}-\left(\gamma^{2}+\frac{1}{4}(\gamma^{2}-1)\left(\frac{1}{|\beta|^{2}}+\Box_{\beta}\right)\right)\mathcal{F}_{2a}^{(2)}(|\beta|)\right]\label{eq:NLOMassiveImpulse} \\
        \Delta p_{1}^{(2),\rho}\big|_{2b}&=\frac{G^{2}m_{1}m_{2}}{4\pi|b|^{2}}|\beta|\beta^{\rho}\left[\frac{\bc_{1,1}^{2}\bc_{2,1}^{2}(m_{1}+m_{2})}{\gamma^{2}-1}\frac{\partial\mathcal{F}^{(1)}(2|\beta|)}{\partial|\beta|}+\frac{1}{2}(\bc_{1,1}^{2}\bc_{2,2}m_{1}+\bc_{2,1}^{2}\bc_{1,2}m_{2})\left(\frac{1}{|\beta|^{2}}+\Box_{\beta}\right)\mathcal{F}^{(2)}_{2a}(|\beta|)\right]\notag \\
        &\quad-\frac{G^{2}m_{1}m_{2}}{4\pi|b|^{2}}\frac{\bc_{1,1}^{2}\bc_{2,1}^{2}(m_{2}\check{v}_{1}^{\rho}-m_{1}\check{v}_{2}^{\rho})}{\sqrt{\gamma^{2}-1}}|\beta|^{2}\left(1+\frac{1}{2}\Box_{\beta}\right)\mathcal{F}^{(2)}_{2b}(|\beta|),\notag
    \end{align}
       where $\Box_{x}\equiv\partial^{2}/\partial x_{\sigma}\partial x^\sigma$.
    Three novel Fourier transforms appear here:
    \begin{equation}\label{eq:OneLoopFTsMomentumSpace}
    \begin{aligned}
        \mathcal{F}^{(2)}_{1}(|\beta|)&=\mu\int_{q}\frac{(-q^{2})^{\varepsilon-\frac{1}{2}}}{(\mu^2-q^{2})^{1+2 \varepsilon}}B_{Q(\mu)}\left(\tfrac{1}{2}-\varepsilon ,\tfrac{1}{2}-\varepsilon \right) \\
        \mathcal{F}^{(2)}_{2a}(|\beta|)&=\frac{\partial}{\partial|b|}\int_{q}\frac{(-q^{2})^{-\frac{3}{2}}}{(4\mu^{2}-q^2)^{\varepsilon}}\sum_{\sigma=\pm}\sigma B_{\frac{1}{2}+\frac{\sigma}{2}\sqrt{Q(2\mu)}}\left(\tfrac{1}{2}-\varepsilon,\tfrac{1}{2}-\varepsilon\right) \\
        \mathcal{F}^{(2)}_{2b}(|\beta|)&=\int_{q}\frac{(-q^{2})^{-\frac{1}{2}}}{(4\mu^{2}-q^2)^{\frac{1}{2}+\varepsilon}}B_{Q(2\mu)}\left(\tfrac{1}{2},-\varepsilon\right),
    \end{aligned}
    \end{equation}
        involving incomplete beta functions $B_{z}(x,y)$ and $Q(\mu)\equiv-q^{2}/(\mu^{2}-q^{2})$.
\end{widetext}

Out of the three transforms above, only $\mathcal{F}^{(2)}_{2a}$ involves a divergence in the limit $\varepsilon\rightarrow0$, and so must be performed in dimensional regularisation.
We keep the dependence on the dimensional regulator in the other two for the sake of unifying the discussion below.

\subsection{Performing the Fourier transforms}

Though they may appear daunting, the transforms which remain to be done can all be brought to parametric integrals over a single variable, facilitating numerical integration.
While the three integrals naturally demand individualised treatment, two crucial steps are involved in the evaluation of all three.

First is the use of the integral representation of the incomplete beta function:
\begin{align}
    B_{z}(x,y)&=\int_{0}^{z}{\rm d}u\,u^{x-1}(1-u)^{y-1}\notag \\
    &=z^{x}\int_{0}^{1}{\rm d}u\,u^{x-1}(1-zu)^{y-1}.\label{eq:BetaIntegralForm}
\end{align}
The change of variables $u\rightarrow zu$ helps to rationalise dependence on the variable being Fourier transformed.
Then, apart from the phase $\exp(iq\cdot b)$, the transfer momentum only enters through its magnitude $|q|^{2}=-q^{2}$.
This means that the Fourier transforms can be recast as Hankel transforms using
\begin{align*}
    \int_{q}\,f(|q|)&=\frac{|b|^{1-\frac{d-2}{2}}}{(2\pi)^{\frac{d-2}{2}}\sqrt{\gamma^{2}-1}} \\
    &\times\int_{0}^{\infty}{\rm d}|q|\,|q|^{\frac{d-2}{2}}f(|q|)J_{\frac{d-2}{2}-1}(|q||b|),
\end{align*}
with $J_{\nu}(z)$ Bessel functions of the first kind.
If the integrand takes the form $f(|q|)=(\Delta-q^{2})^{-\alpha}$ for some power $\alpha$, then the Hankel transform evaluates to
\begin{equation}\label{eq:GeneralHankelTransform}
\begin{aligned}
    \int_{q}\,(\Delta-q^{2})^{-\alpha}&=\frac{|b|^{\alpha-\frac{d-2}{2}}\Delta^{\frac{d}{4}-\frac{1+\alpha}{2}}}{(2\pi)^{\frac{d-2}{2}}\sqrt{\gamma^{2}-1}} \\
    &\times\frac{2^{1-\alpha}}{\Gamma(\alpha)}K_{1+\alpha-\frac{d}{2}}(|b|\sqrt{\Delta}).
\end{aligned}
\end{equation}
This generalises the familiar formula with $\Delta=0$; see e.g. eq.~(B.2) of ref.~\cite{Jakobsen:2021zvh}.

What remains is to bring the integrand into the form demanded by \cref{eq:GeneralHankelTransform}.
For $\mathcal{F}^{(2)}_{1}$ and $\mathcal{F}^{(2)}_{2a}$, this is achievable through the introduction of Feynman parameters (and some rationalisation of roots for the latter), while $\mathcal{F}^{(2)}_{2b}$ is already in this form upon inserting \cref{eq:BetaIntegralForm}.
In all three cases we succeed in performing all but one of the parametric integrals, producing (in the limit $\varepsilon\rightarrow0$)
\begin{equation}\label{eq:OneLoopFTs}
\begin{aligned}
    \mathcal{F}^{(2)}_{1}(|\beta|)&=-\frac{1}{4\pi\sqrt{\gamma^{2}-1}} \\
    &\times\int_{0}^{1}{\rm d}s\,e^{-|\beta|/\sqrt{s}}\frac{\log(1-s)}{s\sqrt{1-s}} \\
    \mathcal{F}^{(2)}_{2a}(|\beta|)&=-\frac{1}{2\pi |\beta|\sqrt{\gamma^{2}-1}} \\
    &\times\left[1-2|\beta|\int_{1}^{\infty}{\rm d}s\,\frac{K_{1}\left(2|\beta|s\right)}{s}\right] \\
    \mathcal{F}^{(2)}_{2b}(|\beta|)&=\frac{1}{2\pi\sqrt{\gamma^{2}-1}}\int_{0}^{1}{\rm d}s\,\frac{K_{0}(\frac{2|\beta|}{\sqrt{1-s}})}{(1-s)\sqrt{s}}.
\end{aligned}
\end{equation}
We see again that the Fourier transforms are indeed only functions of $\mu$ and $|b|$ through their product $|\beta|$, despite their original appearances.
Though these results are already quite compact, we can still go further.

In the case of $\mathcal{F}_{2b}^{(2)}$, a physical consistency condition enables us to perform the final integral, at least in the form it appears in the 2PM impulse.
Specifically, it is not $\mathcal{F}_{2b}^{(2)}$ which is needed, but rather
\begin{align}
    &\left(1+\tfrac{1}{2}\Box_{\beta}\right)\mathcal{F}_{2b}^{(2)}(|\beta|) \\
    &=-\frac{1}{2\pi\sqrt{\gamma^{2}-1}}\int_{0}^{1}{\rm d}s\,K_{0}\left(\tfrac{2|\beta|}{\sqrt{1-s}}\right)\frac{1+s}{(1-s)^{2}\sqrt{s}}.\notag
\end{align}
The final momentum must obey $(p_{1}+\Delta p_{1})^{2}=m_{1}^{2}$, giving the perturbative constraints
\begin{equation}
\begin{aligned}
    p_{1}\cdot\Delta p_{1}^{(1)}&=0 \\
    p_{1}\cdot\Delta p_{1}^{(2)}&=-\frac{1}{2}\left[\Delta p^{(1)}\right]^{2}.
\end{aligned}
\end{equation}
The first of these is easily satisfied by \cref{eq:LOImpulseEvaluated}, thanks to \cref{eq:ImpactParameter,eq:Projector}.
On the other hand, since the term involving $\mathcal{F}_{2b}^{(2)}$ is the only one not orthogonal to the initial momentum, the second constraint is satisfied iff
\begin{align}
    \int_{0}^{1}{\rm d}s\,K_{0}\left(\tfrac{2|\beta|}{\sqrt{1-s}}\right)\frac{1+s}{(1-s)^{2}\sqrt{s}}=[K_{1}(|\beta|)]^{2}.
\end{align}
We have confirmed this equivalence numerically, providing a consistency check on our calculation and evaluating one of the remaining parametric integrals.
In fact, we are also able to perform the final integral in $\mathcal{F}^{(2)}_{2a}$; its evaluated form is, however, less compact, so we leave it to \cref{eq:FT22aEvaluated}.
All in all, our analysis leaves only one parametric integral -- that in $\mathcal{F}_{1}^{(2)}$ -- unevaluated in the linear impulse to NLO.

All three transforms at 2PM reduce to integrals over $K_{\nu}(z)$.\footnote{In the case of $\mathcal{F}^{(2)}_{1}$, we have used $K_{1/2}(z)\propto e^{-z}/\sqrt{z}$.}
These functions exhibit the property that $\lim_{z\rightarrow\infty}z^n K_{\nu}(az)=0$ for any $a\geq0$, $\nu$ and $n$, such that, as expected, the worldlines decouple at 1PM and for families 1 and 2b at 2PM as $|\beta|\rightarrow\infty$ -- i.e.~as the scalar mass grows.
In the case of family 2a, the massive scalar does not mediate long-range effects.
Its mass rather controls the strength of the graviton-worldline coupling.
This contribution to the impulse consequently grows with the scalar mass; we elaborate on this phenomenon in the next section.
Most importantly, this contribution, as well as all others, shrinks with the dimensionless ratio $Gm_{i}/|b|$ controlling the PM expansion.

We have presented here the most compact forms we have found for the Fourier transforms.
Alternate forms are given in \Cref{app:FourierTransforms}.

\section{Limits of the impulse}\label{sec:Limits}

\subsection{The large mass limit}\label{sec:LargeMass}

As we have indicated above, up to 2PM it is only family 2a at NLO which survives in the $|\beta|\rightarrow\infty$ limit.
Diagrammatically, one can understand this as the pinching of the scalar propagators in family 2a establishing a new effective worldline vertex where the scalar ``cloud'' surrounding a worldline has been integrated out.
Let us quantify this effect and give it a physical interpretation.

Consider the triangle subtopology in family 2a in which the graviton line, carrying outgoing momentum $k$, is amputated, and the two scalar lines attach to the static worldline $i$:
\begin{equation}\label{eq:Family2aSubdiagram}
\Lambda_{i}^{\mu\nu}(k)\;=\;
\vcenter{\hbox{
\begin{tikzpicture}[line cap=round,line join=round,baseline=-1.8cm]
        \path [draw=black!40, worldlineStatic] (-5,-.375) -- (-2.7,-.375);
        \path [draw=black, photon] (-4,-.875) -- (-4, -1.525);
        \draw [fill] (-4,-1.525) circle (0) node [below] {$\mu,\nu$};
        \node at (-3.7,-1.2) {$\,\,\downarrow k$};
        \fill [draw=black] (-4,-0.875) circle (1.75pt);
        \path [draw=black, cscalar2] (-4.5,-.375) -- (-4, -0.875);
        \node at (-4.65,-0.75) {$q_{1}\,\rotatebox{-45}{$\rightarrow$}$};
        \fill [draw=black] (-4.5,-.375) circle (1.75pt);
        \path [draw=black, cscalar2] (-3.5,-.375) -- (-4, -0.875);
        \node at (-3.35,-0.75) {$\rotatebox{225}{$\rightarrow$}\,q_{2}$};
        \fill [draw=black] (-3.5,-.375) circle (1.75pt);
\end{tikzpicture}}}\,.
\end{equation}
Inserting Feynman rules and tensor reducing to our master integrals in \Cref{app:LoopIntegrals}, one finds the manifestly transverse result
\begin{align}\label{eq:EffectiveVertexExact}
    \Lambda_{i}^{\mu\nu}(k)&=\frac{i\kappa\,c_{i,1}^{2}}{4}\,e^{ik\cdot b_{i}}\,\dd(k\cdot v_{i})
    \Big[a(k)\,v_{i}^{\mu}v_{i}^{\nu}\notag \\
    &\qquad\quad+b(k)\Big(\eta^{\mu\nu}-v_{i}^{\mu}v_{i}^{\nu}-\frac{k^{\mu}k^{\nu}}{k^{2}}\Big)\Big],\notag \\
    a(k)&=J_{001}-\frac{k^{2}}{2}J_{011}\,, \\
    b(k)&=\frac{1}{2}J_{001}-\Big(\mu^{2}+\frac{k^{2}}{4}\Big)J_{011}\,,\notag
\end{align}
satisfying the graviton Ward identity $k_{\mu}\Lambda_{i}^{\mu\nu}=0$ exactly.
In momentum space, the $|\beta|\rightarrow\infty$ limit corresponds to $|k|\ll\mu$.\footnote{To see this, note that the graviton probes distances $|k|\sim1/|b|\ll\mu$, with the inequality following from $|\beta|=\mu|b|\gg1$ as $|\beta|\rightarrow\infty$.}
Inserting the master integrals at $\varepsilon=0$ (i.e.~in four spacetime dimensions) 
we have $J_{001}=\mu/4\pi$, while in this limit $J_{011}\rightarrow1/(8\pi\mu)$, so that $b(k)$ is suppressed relative to $a(k)\rightarrow\mu/4\pi$.
Consequently,
\begin{align}
    \Lambda_{i}^{\mu\nu}(k)\overset{|\beta|\rightarrow\infty}{\longrightarrow}\frac{i\kappa\,c_{i,1}^{2}\mu}{16\pi}\,v_{i}^{\mu}v_{i}^{\nu}\,e^{ik\cdot b_{i}}\,\dd(k\cdot v_{i}).
\end{align}
This is nothing but the standard graviton--worldline mass monopole vertex (see e.g.~\cite{Mogull:2020sak})
\begin{equation}\label{eq:EffectiveVertexRule}
\begin{aligned}
    \vcenter{\hbox{
    \begin{tikzpicture}[line cap=round,line join=round,baseline=-0.4cm]
        \path [draw=black!40, worldlineStatic] (0,0) -- (1.6,0);
        \path [draw=black, photon] (0.8,-0.1) -- (0.8, -1.) node [midway, right] {$\downarrow k$};
        \draw [fill] (0.8,-1.) circle (0) node [below] {$\mu,\nu$};
        \draw [fill=white, draw=black] (0.8,0) circle (3pt);
        \node[cross=2.2pt] at (0.8,0) {};
    \end{tikzpicture}}}\,\,=-\frac{i\kappa}{2}\,\delta m_{i}\,v_{i}^{\mu}v_{i}^{\nu}
    e^{ik\cdot b_{i}}\,\dd(k\cdot v_{i})\,,
\end{aligned}
\end{equation}
carrying the effective mass
\begin{equation}\label{eq:DeltaM}
    \delta m_{i}=-\frac{c_{i,1}^{2}\,\mu}{8\pi} < 0\,.
\end{equation}
We have used a cross to indicate the effective vertex encoding $\lim_{|\beta|\rightarrow\infty}\Lambda_{i}^{\mu\nu}(k)$.
Crucially, $\delta m_{i}$ is negative for any sign of the
scalar charges, such that the scalar {``cloud’’} invariably screens the gravitational mass of the compact object, reducing it to
\begin{align}\label{eq:EffectiveMass}
    m_{i}^{\rm eff}=m_{i}+\delta m_{i}
    =m_{i}\left(1-\frac{\bc_{i,1}^{2}\,G m_{i}\mu}{8\pi}\right),
\end{align}
where we inserted the scaling \cref{eq:CoefficientScaling2}. We now see that
the magnitude of the screening is controlled by the ratio of the Schwarzschild radius
of the compact object to the range of the scalar interaction, $\lambda_{i}\equiv 2Gm_{i}\mu$,
assuming naturalness for the scalar charge $\bar{c}_{i,1}$. In fact, $\lambda_{i}$ may take arbitrary
values without leaving the post-Minkowskian perturbative domain $|b|\gg Gm_{i}$.

As a check, we take the $|\beta|\rightarrow\infty$ limit of the family 2a contribution in \cref{eq:NLOMassiveImpulse}:
\begin{equation}\label{eq:Family2aAsymptotics}
\begin{aligned}
    \Delta p_{1}^{(2),\rho}\big|_{2a}
    &\overset{|\beta|\rightarrow\infty}{\longrightarrow}\frac{G^{2}m_{1}m_{2}}{|b|^{2}}\beta^{\rho} \\
    &\quad\times\frac{\left(\bc_{1,1}^{2}m_{1}+\bc_{2,1}^{2}m_{2}\right)
    (2\gamma^{2}-1)}{4\pi\sqrt{\gamma^{2}-1}}\,,
\end{aligned}
\end{equation}
all other contributions up to 2PM vanishing.
This is precisely the result of inserting $m_{i}\rightarrow m_{i}^{\rm eff}$ into the 1PM gravitational impulse and extracting the correction generated by the mass shift.
\subsection{The massless limit}

Predictably, the linear impulse simplifies appreciably in the opposite limit, i.e. when $|\beta|\rightarrow0$.
At LO, this limit of \cref{eq:LOImpulse} is smooth and gives
\begin{align}
    \lim_{|\beta|\rightarrow0}\Delta p_{1}^{(1),\rho}&=-\frac{G m_{1}m_{2 }\bc_{1,1}\bc_{2,1}}{2\pi\sqrt{\gamma^{2}-1}}\frac{b^\rho}{|b|^2}.
\end{align}
At NLO, we must be cautious about whether or not this limit commutes with the remaining parametric integrals; in the forms of \cref{eq:OneLoopFTs}, it does not.\footnote{For example, the argument of $K_{1}(2|\beta|s)$ is not small as long as $(2|\beta|)^{-1}\lesssim s$, which lies in the integration domain when $|\beta|\rightarrow0$.}

Despite this non-commutativity of the limits, we have established (numerically for the first, analytically for the second and third) that the Fourier transforms possess the asymptotic forms
\begin{align}\label{eq:FTMasslessLimits}
    \mathcal{F}^{(2)}_{1}(|\beta|)&\overset{|\beta|\rightarrow0}{\longrightarrow}\frac{\pi-4|\beta|}{8\sqrt{\gamma^{2}-1}} \\
    \mathcal{F}^{(2)}_{2a}(|\beta|)&\overset{|\beta|\rightarrow0}{\longrightarrow}
    -\frac{1}{2\sqrt{\gamma^{2}-1}}
     \\
    \left(1+\tfrac{1}{2}\Box_{\beta}\right)\mathcal{F}^{(2)}_{2b}(|\beta|)&\overset{|\beta|\rightarrow0}{\longrightarrow}-\frac{1}{2\pi\sqrt{\gamma^{2}-1}|\beta|^{2}}.
\end{align}
In the first line, the leading term as $|\beta|\rightarrow0$ drops out from the impulse after taking the required derivatives.
In the massless limit, the impulse becomes
\begin{align}
    \lim_{|\beta|\rightarrow0}\Delta p_{1}^{(2),\rho}\Big|_{1}&=-\frac{G^{2}m_{1}m_{2}\bc_{1,1}\bc_{2,1}(m_{1}+m_{2})}{2\sqrt{\gamma^{2}-1}}\frac{b^{\rho}}{|b|^{3}}\notag \\
    \lim_{|\beta|\rightarrow0}\Delta p_{1}^{(2),\rho}\Big|_{2a}&=\frac{G^{2}m_{1}m_{2}(\bc_{1,1}^{2}m_{1}+\bc_{2,1}^{2}m_{2})}{16(\gamma^{2}-1)^{-1/2}}\frac{b^{\rho}}{|b|^{3}}\label{eq:NLOMasslessImpulse} \\
    \lim_{|\beta|\rightarrow0}\Delta p_{1}^{(2),\rho}\Big|_{2b}&=\frac{G^{2}m_{1}m_{2}\bc_{1,1}^{2}\bc_{2,1}^{2}(m_{2}\Check{v}_{1}^{\rho}-m_{1}\Check{v}_{2}^{\rho})}{8\pi^{2}(\gamma^{2}-1)|b|^{2}}\notag \\
    &-\frac{G^{2}m_{1}m_{2}(\bc_{1,1}^{2}\bc_{2,2}m_{1}+\bc_{2,1}^{2}\bc_{1,2}m_{2})}{16\pi\sqrt{\gamma^{2}-1}}\frac{b^{\rho}}{|b|^{3}}.\notag
\end{align}
We have confirmed that this coincides with setting $\mu=0$ from the start, such that the massless limit is smooth at NLO as well.

Crucially, despite the smoothness of the massless limit, it is not possible to build up the massive result by resumming an expansion in small $\mu$, where the Fourier transforms are analytically tractable.
The reason for this is that the transforms do not commute with such a series approach.
In fact, this is already apparent from the momentum space expression of the 1PM Fourier transform, \cref{eq:TreeLevelFTMomentumSpace}.
For any small but non-zero value of $\mu$, this integral has a region where $0\leq-q^2\lesssim \mu^{2}$ in which a small-mass expansion is not justified.
Indeed, attempting to solve this integral through a series expansion produces a modified Bessel function of the first kind, in contradistinction to \cref{eq:TreeLevelFT}.

\section{The scattering angle to 2PM}\label{sec:ScatteringAngle}

In the conservative case, the entire information carried by the impulse is encapsulated by the scattering angle, which may be defined by \cite{Jakobsen:2023oow}
\begin{align}\label{eq:ScatteringAngleDefinition}
    \sin\chi =\frac{\check b\cdot\Delta p_{1}}{p_{\infty}},
\end{align}
where $\check b^{\mu}= b^{\mu}/|b|$ and $p_{\infty}$ is the centre-of-mass momentum,
\begin{align}
    p_{\infty}=m_{1}m_{2}\sqrt{\frac{\gamma^{2}-1}{(p_{1}+p_{2})^{2}}}\,.
\end{align}
The component of the impulse along the $\check v_{i}^{\mu}$ directions follows from the conservation
equations.
This definition accounts for the attractive or repulsive quality of the encounter.
\Cref{eq:ScatteringAngleDefinition} implies that,
up to 2PM precision, the scattering angle simply follows from the $b^{\mu}$ components of 
eqs.~(\ref{eq:LOImpulseEvaluated}) and (\ref{eq:NLOMassiveImpulse}).

\begin{figure*}[t]
\centering

\begin{minipage}{0.4\textwidth}
\centering
\includegraphics[scale = 0.25, trim=2.5cm 0cm 0cm 0cm,]{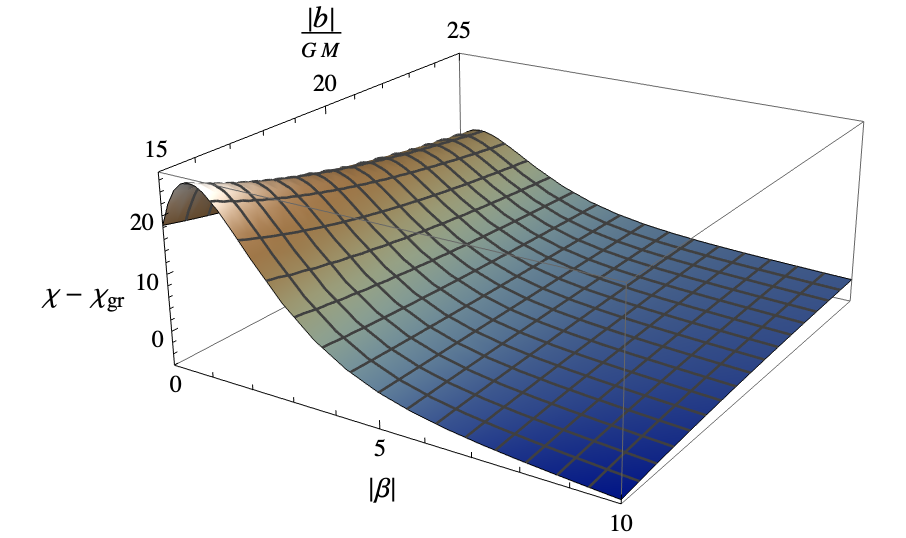}
\end{minipage}%
\qquad\qquad
\begin{minipage}{0.4\textwidth}
\centering
\includegraphics[scale = 0.25, trim=1.5cm 0cm 0cm 0cm,]{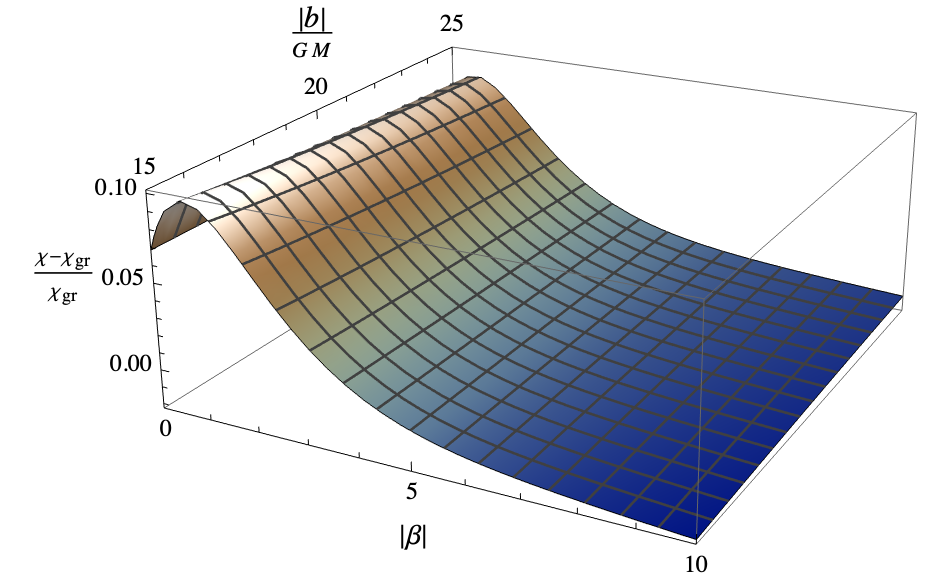}
\end{minipage}

\caption{\label{fig:ScatteringAngleAttract} The scattering angle as a function of the magnitude of the impact parameter $|b|$ and the scalar's mass parameter $|\beta|=\mu |b|$ for equal masses, $v/c=1/5$, and $\bar{c}_{i,j}=1$ (\emph{attractive} case).
\textit{Left}: The difference (in degrees) between the total and the purely gravitational scattering angle up to 2PM order.
\textit{Right}: The size of the scalar contribution to the scattering angle relative to the gravitational one. }

\end{figure*}

Note that \cref{eq:ScatteringAngleDefinition} is linear in the impulse, so the total scattering angle up to 2PM is simply given by the sum
\begin{align}
    \chi=\chi_{\rm gr}+\chi_{\rm sc}+\cO(G^3),
\end{align}
where $\chi_{\rm sc}$ is the scattering angle due to the scalar exchange contributions evaluated above.
The purely gravitational angle up to 2PM takes the form
\begin{equation}
\begin{aligned}
    \chi_{\rm gr}&=\frac{2\sqrt{(p_{1}+p_{2})^{2}}}{\gamma^{2}-1}\left[\frac{G}{|b|}(2\gamma^{2}-1)\right. \\
    &\quad\left.+\frac{3\pi G^{2}}{8|b|^{2}}(m_{1}+m_{2})(5\gamma^{2}-1)\right],
\end{aligned}
\end{equation}
as was first derived in refs.~\cite{Westpfahl:1985tsl,Damour:2017zjx}, and can be extracted from the linear impulse computed to 2PM in ref.~\cite{Mogull:2020sak}.

\begin{figure*}[t]
\centering

\begin{minipage}{0.4\textwidth}
\centering
\includegraphics[scale = 0.25]{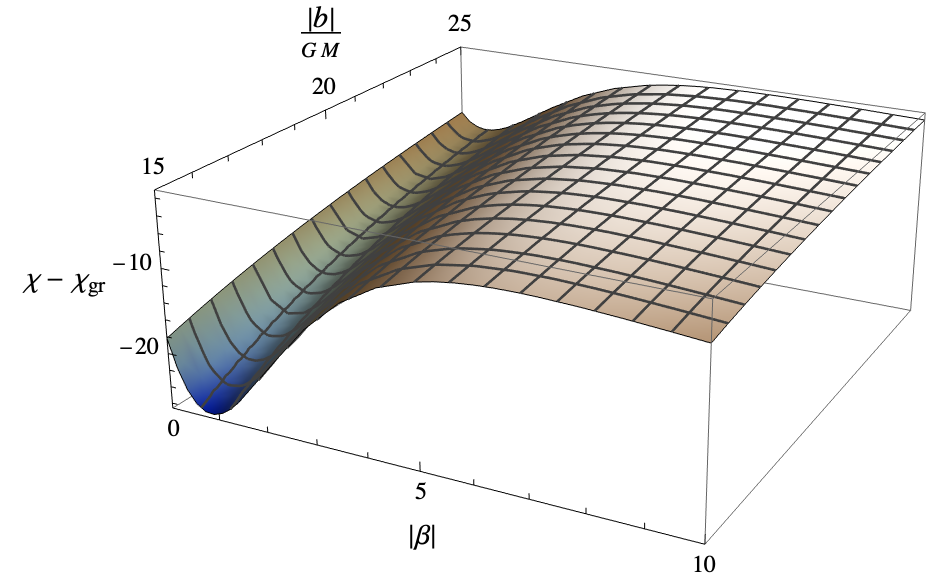}
\end{minipage}%
\qquad\qquad
\begin{minipage}{0.4\textwidth}
\centering
\includegraphics[scale = 0.25]{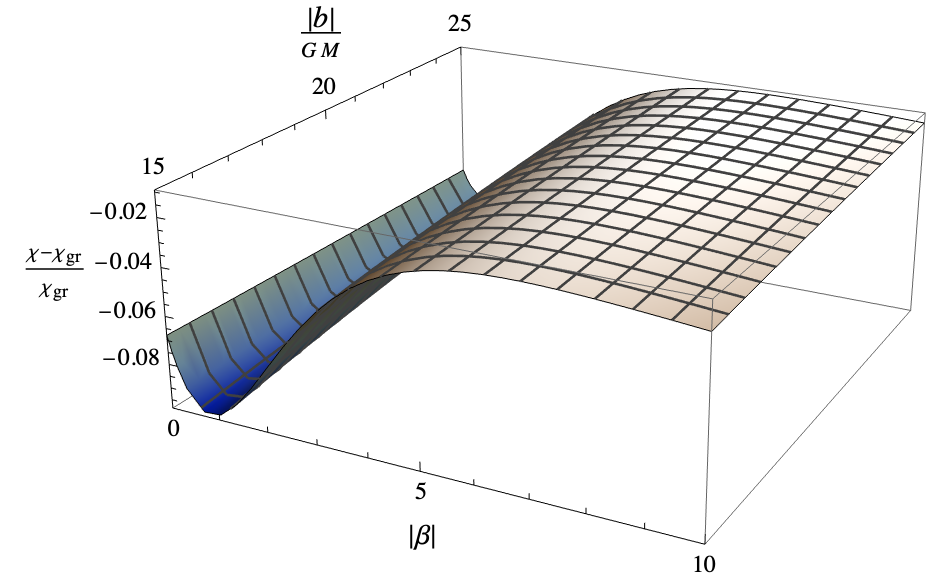}
\end{minipage}

\caption{\label{fig:ScatteringAngleRepulse}  Same as FIG.~\ref{fig:ScatteringAngleAttract} for the \emph{repulsive} case with $\bar{c}_{2,1}=-1$ and all other $\bar{c}_{i,j}=1$.
\textit{Left}: The difference (in degrees) between the total and the purely gravitational scattering angle up to 2PM order.
\textit{Right}: The size of the scalar contribution to the scattering angle relative to the gravitational one.
}

\end{figure*}

To exemplify the effect of the scalar mediation, we focus on the equal mass case $m_{1}=m_{2}=M/2$ for $M=m_{1}+m_{2}$.
Then, the scalar coupling constants $c_{i,j}$ can be taken to be constrained in their magnitude thanks to \cref{eq:CoefficientScaling2} and Wilsonian naturalness demanding that the dimensionless $\bar{c}_{i,j}$ be near unity.
This leaves only two parameters which control the scattering: $|\beta|=\mu|b|$, probing the scalar mass, and $GM/|b|$, measuring the perturbative strength of the gravitational attraction.
In this setting, we visualise the effect of including the scalar mediator in \cref{fig:ScatteringAngleAttract,fig:ScatteringAngleRepulse}.
There, the difference between the total scattering angle $\chi$ and the purely gravitational angle $\chi_{\rm gr}$ is plotted as a function of both free parameters, as well as the ratio of this difference to the purely gravitational scattering angle.
We set $\gamma=1.02062$ in all cases, corresponding to a relative velocity $v/c=1/5$, {and plot in the range $|b|/GM>15$ which is well within the perturbative PM region (see fig.~4 of \cite{Driesse:2024feo} for
a state-of-the-art PM comparison to numerical relativity)}.

In these figures, we illustrate two different combinations of the dimensionless scalar-worldline couplings.
First, \cref{fig:ScatteringAngleAttract} considers all $\bar{c}_{i,j}=1$.
Consequently, the interaction is attractive, as is reflected in the sign of the scattering angle being the same as the gravitational one.
Second, \cref{fig:ScatteringAngleRepulse} flips the sign on $\bar{c}_{2,1}$, making the scalar force repulsive.

\begin{figure}[t]
\centering

\includegraphics[width = 0.47\textwidth]{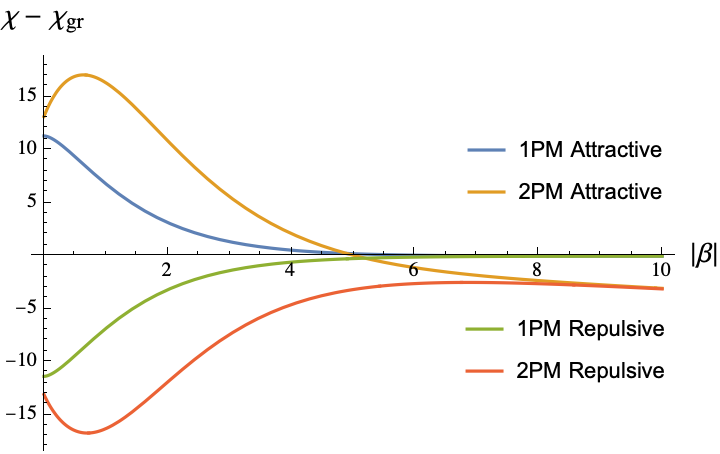}

\caption{\label{fig:Resonance} The resonance effect in the scattering angle
at $|\beta|=\mu|b|\sim 1$
due to the NLO correction in the attractive and repulsive scenarios of figs.~\ref{fig:ScatteringAngleAttract} and \ref{fig:ScatteringAngleRepulse},
plotted at $|b|/GM=20$ and $v/c=1/5$.
The unbounded repulsive falloff at large $|\beta|$, common to both scenarios,
reflects the screening of the worldline masses by the scalar self-field cloud,
$\delta m_{i}=-c_{i,1}^{2}\mu/8\pi$; see \cref{sec:LargeMass}.}

\end{figure}

On top of the expected decay of the scattering angle as $|b|/GM\rightarrow\infty$, the inclusion of the massive mediator produces a qualitative deviation from the purely gravitational dynamics, in the form of {resonance-like behavior} when $\mu\sim1/|b|$.
At this peak, the scalar contribution can reach as much as one-tenth the value of the gravitational one.
It is noteworthy that this extremum only emerges once NLO effects are incorporated, a fact reflecting the monotonicity in $|\beta|$ of the LO result of \cref{eq:LOImpulseEvaluated}; see figure~\ref{fig:Resonance}. 

In the repulsive case, a maximum is attained by $\chi_{\rm sc}$ around $\mu\sim 6.8/|b|$, while attractive dynamics do not exhibit a second extremum.
At large $|\beta|$, both attractive and repulsive configurations grow indefinitely in magnitude due to the dominance of the graviton-scalar coupling captured by family 2a above.
From our analysis in \Cref{sec:LargeMass}, we understand this growth at large $|\beta|$ to not be a genuine long-range repulsive force, but rather a finite screening of the worldline masses by self-interactions through the scalar field.

\section{Conclusions}\label{sec:Conclusions}

We have introduced a scalar field into the WQFT action for a spinless gravitating particle, producing a simple model of effects such as dark matter or exotic compact celestial objects.
New couplings between the scalar and the worldline lead to subtleties in organising the PM expansion, and the inverse-length scale $\mu$ associated with the scalar's mass complicates integration at subleading orders.
In this work, we have sought to bring both of these aspects under analytic control, in order to evaluate their imprint on compact-binary scattering.

{Considering the scales in the classical physics problem, we concluded that, 
upon invoking naturalness, the expansions in gravitational and scalar-worldline couplings are not independent.
Rather, both types of couplings count equally at a given PM order.}

With this perspective, we evaluated the contribution from the mediating scalar to the linear impulse induced in a two-body scattering encounter, up to NLO (i.e.~2PM).
The equal relevance of both types of worldline couplings led to a division of the 2PM impulse into three distinct contributions, roughly corresponding to long-range scalar exchange, an effective graviton-worldline coupling, and tidal effects.
Completing the impulse required performing Fourier transforms with complicated dependence on the scalar's mass.
We outlined a method for doing so, which resulted in compact expressions involving at most a single unevaluated parametric integral over modified Bessel functions of the second kind.

Studying the regime of a large scalar mass, we uncovered a screening mechanism: 
{integrating out the scalar sourced by each compact object reduces the monopole coupling to gravity by $\delta m_{i}=-c_{i,1}^{2}\mu/8\pi$.}
It will be interesting to understand whether this is an observable effect, or rather amounts to a finite renormalisation and is absorbed upon expressing 
the coupling of the graviton to the worldline in
 terms of the effective masses $m_{i}^{\rm eff}$.
 
Finally, the scattering angle is a useful quantity for understanding the numerical significance of this new interaction channel.
We probed the parameter space spanned by the scalar's mass for equal worldline masses, and established the importance of the scalar exchange relative to a purely gravitational interaction.
While our analysis does not allow us to assess observability, the presence of a 
{maximum} in the scattering angle at $\mu\sim1/|b|$, reaching up to one-tenth of the gravitational contribution, makes clear that a massive scalar mediator could significantly modify binary scattering dynamics; see also ref.~\cite{Evstafyeva:2024qvp}.
At least in principle, such a {peak} could serve as a direct indicator of a non-zero scalar mass in scattering observations: sampling over impact parameters would single out the scale $|b|\sim1/\mu$.


{Looking ahead, it will be interesting to investigate whether the NLO maximum we observed in the present model finds analogs in more astrophysically relevant theories, such as the effective description of solitonic boson stars of ref.~\cite{Damour:2025oys}.
In the event that we observe such a phenomenon, building on the comparison to numerical relativity performed there would be crucial to assessing whether this NLO phenomenon is indeed observable.}





\subsection*{Data availability}

We provide a computational notebook with machine-readable results of this work
in the zenodo.org repository submission \cite{zenodo}.

\subsection*{Acknowledgements}

We are grateful to Lara Bohnenblust, Mathias Driesse, Jitze Hoogeveen, Gustav Jakobsen, Rob Klabbers, Johannes Pirsch, Benjamin Sauer, Kathrin Stoldt and Johann Usovitsch for helpful conversations. J.P.~thanks Oluwadamilola Babayemi and Gustav Mogull for earlier collaborations on this topic.
This work was funded by the European Union through the
European Research Council under grant ERC Advanced Grant 101097219 (GraWFTy).
Views and opinions expressed are however those of the authors only and do not necessarily reflect those of the European Union or European Research Council Executive Agency. Neither the European Union nor the granting authority can be held responsible for them.

\newpage
\appendix

\begin{widetext}

\section{Feynman rules}\label{app:FeynmanRules}

We list here the Feynman rules derived from the actions \cref{eq:BulkAction,eq:WQFTAction2} needed for the calculations presented in the main body of the paper.
The three single-scalar-worldline vertices are
\begin{align}
    \begin{tikzpicture}[line cap=round,line join=round,baseline=-0.4cm]
        \path [draw=black!40, worldlineStatic] (0,0) -- (1.6,0);
        \path [draw=black, cscalar2] (0.8,0) -- (0.8, -1.) node [midway, right] {$\downarrow\ell$};
        \fill [draw=black] (0.8,0) circle (1.75pt);
    \end{tikzpicture}
    &=i e^{i\ell\cdot b_i} \dd\!\left(\ell\cdot v_i\right)c_{i,1} \\
    \begin{tikzpicture}[line cap=round,line join=round,baseline=-0.4cm]
        \path [draw=black!40, worldlineStatic] (0,0) -- (0.8,0);
        \path [draw=black, zParticle2] (0.8,0) -- (1.6,0) node [midway, above] {$\omega_{1}$};
        \path [draw=black, cscalar2] (0.8,0) -- (0.8, -1.) node [midway, right] {$\downarrow\ell$};
        \fill [draw=black] (0.8,0) circle (1.75pt);
        \draw [fill] (1.6,0) circle (0) node [right] {$\rho$};
    \end{tikzpicture}
    &=-e^{i\ell\cdot b_{i}}\dd(\omega_{1}+\ell\cdot v_{i})c_{i,1}\ell^{\rho} \\
    \begin{tikzpicture}[line cap=round,line join=round,baseline=-0.4cm]
        \path [draw=black, zParticle2] (0,0) -- (0.8,0);
        \path [draw=black, zParticle2] (0.8,0) -- (1.6,0) node [midway, above] {$\omega_{1}$};
        \path [draw=black, cscalar2] (0.8,0) -- (0.8, -1.) node [midway, right] {$\downarrow\ell$};
        \fill [draw=black] (0.8,0) circle (1.75pt);
        \node at (0.15,0.25) {$\leftarrow\omega_{2}$};
        \draw [fill] (0,0) circle (0) node [left] {$\nu$};
        \draw [fill] (1.6,0) circle (0) node [right] {$\rho$};
    \end{tikzpicture}
    &=-i e^{i\ell\cdot b_i}\dd(\omega_{1}+\omega_{2}+\ell\cdot v_{i})c_{i,1}\ell^{\nu}\ell^{\rho}.
\end{align}
Arrows on the worldline perturbations indicate the direction of causality flow, while the arrow pointing left in the third rule denotes that the energy is flowing out of the vertex.

When two scalar lines are involved, the rules needed are
\begin{align} \label{eq:A4}
    \begin{tikzpicture}[line cap=round,line join=round,baseline=-0.4cm]
        \path [draw=black!40, worldlineStatic] (0,0) -- (1.6,0);
        \path [draw=black, cscalar2] (0.8,0) -- (0.4, -1) node [midway, left] {$\ell_{1}\rotatebox{68}{$\leftarrow$}$};
        \path [draw=black, cscalar2] (0.8,0) -- (1.2, -1) node [midway, right] {$\rotatebox{113}{$\leftarrow$}\ell_{2}$};
        \fill [draw=black] (0.8,0) circle (1.75pt);
    \end{tikzpicture}
    &=2ie^{i(\ell_{1}+\ell_{2})\cdot b_{i}}\dd(\ell_{1}\cdot v_{i}+\ell_{2}\cdot v_{i})c_{i,2} \\
    \begin{tikzpicture}[line cap=round,line join=round,baseline=-0.4cm]
        \path [draw=black!40, worldlineStatic] (0,0) -- (0.8,0);
        \path [draw=black, zParticle2] (0.8,0) -- (1.6,0) node [midway, above] {$\omega$};
        \path [draw=black, cscalar2] (0.8,0) -- (0.4, -1) node [midway, left] {$\ell_{1}\rotatebox{68}{$\leftarrow$}$};
        \path [draw=black, cscalar2] (0.8,0) -- (1.2, -1) node [midway, right] {$\rotatebox{113}{$\leftarrow$}\ell_{2}$};
        \fill [draw=black] (0.8,0) circle (1.75pt);
        \draw [fill] (1.6,0) circle (0) node [right] {$\rho$};
    \end{tikzpicture}
    &=-2e^{i(\ell_{1}+\ell_{2})\cdot b_{i}}\dd(\omega+\ell_{1}\cdot v_{i}+\ell_{2}\cdot v_{i})c_{i,2}(\ell_{1}^{\rho}+\ell_{2}^{\rho})\,.
\end{align}
Finally, one vertex is needed in the bulk:
\begin{align}
    \begin{tikzpicture}[line cap=round,line join=round,baseline=-0.2cm]
        \coordinate (origin) at (0,0);
        \coordinate (tL) at (-0.707,0.707);
        \coordinate (tR) at (0.707,0.707);
        \coordinate (b) at (0,-1);
        \path [draw=black, cscalar2] (b) -- (origin) node [midway, right] {$\downarrow \ell_{1}$};
        \path [draw=black, gravitonPlain] (tR) -- (origin);
        \path [draw=black, cscalar2] (tL) -- (origin) node [midway, left] {$\ell_{2}\nwarrow$};
        \fill [draw=black] (origin) circle (1.75pt);
        \draw [fill] (tR) circle (0) node [above] {$\nu,\rho$};
    \end{tikzpicture}
    =\frac{i\kappa}{2}\left[\ell_1^\rho\ell_2^\nu +\ell_1^\nu\ell_2^\rho-\eta^{\rho\nu} \left( \mu^2+\ell_1\cdot\ell_2\right)\right].
\end{align}
Rules governing graviton-worldline interactions can be found in e.g. ref.~\cite{Jakobsen:2021zvh}.

\section{Loop integrals}\label{app:LoopIntegrals}

Two integral families arise in the NLO scattering mediated by gravitons and massive scalars.
Evaluated in dimensional regularisation with $d=4-2\varepsilon$, these are
\begin{align}
    I_{n_{1}n_{2}n_{3}}^{(1,2),\pm}=\tilde{\mu}^{2\varepsilon}\int\frac{{\rm d}^{d}\ell}{(2\pi)^{d}}\frac{\dd(v_{1,2}\cdot \ell)}{(v_{2,1}\cdot\ell\pm i0^+)^{n_{1}}D_{1,0}^{n_{2}}D_{2,\mu}^{n_{3}}},\label{eq:IntegralFamily1} \\
    J_{n_{1}n_{2}n_{3}}^{(1,2),\pm}=\tilde{\mu}^{2\varepsilon}\int\frac{{\rm d}^{d}\ell}{(2\pi)^{d}}\frac{\dd(v_{1,2}\cdot \ell)}{(v_{2,1}\cdot\ell\pm i0^+)^{n_{1}}D_{1,\mu}^{n_{2}}D_{2,\mu}^{n_{3}}},\label{eq:IntegralFamily2}
\end{align}
where
\begin{align}
    D_{1,\mu}&=\ell^{2}-\mu^{2}+i0^+ \\
    D_{2,\mu}&=(\ell-q)^{2}-\mu^{2}+i0^+.
\end{align}
While each family can be IBP-reduced to a set of four master integrals, some of these master integrals overlap and some others do not arise in the NLO calculation.
In all, four master integrals are needed, all of which can be performed by introducing Feynman/Schwinger parameters:
\begin{align}
    I_{001}^{(i),\pm}=J_{001}^{(i),\pm}&=-\tilde{\mu}^{2\varepsilon}\frac{\mu^{1-2\varepsilon}}{(4\pi)^{\frac{3}{2}-\varepsilon}}\Gamma\left(-\tfrac{1}{2}+\varepsilon\right) \\
    I_{011}^{(i),\pm}&=\tilde{\mu}^{2\varepsilon}\frac{\Gamma(\frac{1}{2}+\varepsilon)}{(4\pi)^{\frac{3}{2}-\varepsilon}}\frac{(-q^{2})^{-\frac{1}{2}+\varepsilon}}{(\mu^{2}-q^{2})^{2\varepsilon}}B_{\frac{-q^{2}}{\mu^{2}-q^{2}}}(\tfrac{1}{2}-\varepsilon,\tfrac{1}{2}-\varepsilon) \\
    J_{011}^{(i),\pm}&=\frac{\tilde{\mu}^{2\varepsilon}\pi\sec(\pi\varepsilon)(4\mu^{2}-q^{2})^{-\varepsilon}}{(4\pi)^{\frac{3}{2}-\varepsilon}\Gamma\left(\frac{1}{2}-\varepsilon\right)(-q^{2})^{\frac{1}{2}}}\sum_{\sigma=\pm}\sigma B_{\frac{1}{2}+\frac{\sigma\sqrt{-q^{2}}}{2\sqrt{4\mu^{2}-q^{2}}}}\left(\tfrac{1}{2}-\varepsilon,\tfrac{1}{2}-\varepsilon\right) \\
    J_{111}^{(i),\pm}&=\mp i\tilde{\mu}^{2\varepsilon}\frac{4^{2\varepsilon-1}\pi^{\varepsilon-1}}{\sqrt{\gamma^{2}-1}}\frac{\Gamma(1+\varepsilon)(-q^{2})^{-\frac{1}{2}}}{(4\mu^{2}-q^{2})^{\frac{1}{2}+\varepsilon}}B_{\frac{-q^{2}}{4\mu^{2}-q^{2}}}\left(\tfrac{1}{2},-\varepsilon\right),
\end{align}
where $B_{z}(a,b)$ is the incomplete beta function.
The integral $J_{110}^{(i)}$ also appears in the momentum-space result, but in a $q$-independent contribution which becomes ultra-local, and thus irrelevant, in impact-parameter space.

\section{Fourier transforms}\label{app:FourierTransforms}

Here we provide alternative forms of the Fourier transforms needed at 2PM.

\subsection{Alternative forms in terms of Meijer $G$-functions}

The two parametric integrals left over in \cref{eq:OneLoopFTs} can be assigned expressions in terms of limits/derivatives of Meijer $G$-functions.
First, we express the one appearing in $\mathcal{F}^{(2)}_{1}$ as a derivative with respect to an auxiliary variable,
\begin{align}
    \int_{0}^{1}{\rm d}s\,e^{-|\beta|/\sqrt{s}}\frac{\log(1-s)}{s\sqrt{1-s}}&=F^{\prime}(-1/2),
\end{align}
where
\begin{align}
    F(\lambda)\equiv\int_{0}^{1}{\rm d}s\,e^{-|\beta|/\sqrt{s}}\frac{(1-s)^{\lambda}}{s}=\frac{\Gamma(1+\lambda)}{\sqrt{\pi}}G^{3\,0}_{1\,0}\left(\frac{|\beta|^{2}}{4}\left|\,
        \begin{matrix}
             & 1+\lambda & \\
            0 & 0 & \frac{1}{2}
        \end{matrix}
    \right.\right).
\end{align}
In the case of $\mathcal{F}^{(2)}_{2a}$, the indefinite form of the parametric integral is
\begin{align}
    \int{\rm d}s\,\frac{K_{1}\left(2|\beta|s\right)}{s}&=\frac{1}{4}G^{2\,1}_{0\,1}\left(|\beta|s,\frac{1}{2}\left|\,
        \begin{matrix}
            & 1 & \\
            -\frac{1}{2} & \frac{1}{2} & 0
        \end{matrix}
    \right.\right).
\end{align}
We note that ref.~\cite{Bhattacharyya:2024aeq} also encountered this class of functions in their evaluations of related integrals.

Though it is satisfying that all integrals can be expressed in terms of such generalised special functions, these forms are unwieldy.
Indeed, we were not able to take the required limits in order to completely evaluate these integrals.
At this stage, then, these forms do not offer advantages over \cref{eq:OneLoopFTs}.

\subsection{Alternative forms for a smooth massless limit}



As highlighted in the main text, care must be taken to evaluate the massless -- $|\beta|\rightarrow0$ -- limit of the impulse when this limit does not commute with the remaining parametric integrals.
Here we provide forms of the Fourier transforms which are less compact than those in \cref{eq:OneLoopFTs}, but whose massless limit is more easily taken.
We were able to completely evaluate $\mathcal{F}_{2b}^{(2)}$, so here we focus only on the other two transforms at NLO.

Beginning with $\mathcal{F}_{1}^{(2)}$, its second derivative may be recast as
\begin{align}\label{eq:BoxFT21}
    \square_{\beta} \mathcal{F}_1^{(2)}(|\beta|)&=\frac{1}{\pi|\beta|\sqrt{\gamma^2-1}}\left[\frac{\pi}{2}+ \frac{\pi^2}{4} |\beta|\left( \boldsymbol{L}_0(|\beta|)- I_0(|\beta|)\right)- |\beta|\int_{0}^{1}{\rm d}s\,\frac{K_0(|\beta|)-s^2 K_0(|\beta| s)}{1-s^2}\right],
\end{align}
where $\boldsymbol{L}_{\nu}(z)$ is the Struve L-function and $I_{\nu}(z)$ is the modified Bessel function of the first kind.
Now there is no obstruction to commuting the massless limit and the remaining parametric integral, since for small $|\beta|$ there is no region of integration where the arguments of the $K_{0}(z)$ are not small.
In fact, only the first term in square brackets survives the massless limit, such that
\begin{align}
    \square_{\beta} \mathcal{F}_1^{(2)}(|\beta|)\overset{|\beta|\rightarrow0}{\longrightarrow}\frac{1}{2|\beta|\sqrt{\gamma^{2}-1}}\,,
\end{align}
consistent with \cref{eq:FTMasslessLimits}.

Moving to the Fourier transform for family 2a, the parametric integral can be integrated by parts, producing
\begin{equation}\label{eq:FT22aEvaluated}
\begin{aligned}
    \mathcal{F}^{(2)}_{2a}(|\beta|)&=\frac{|\beta|}{\sqrt{\gamma^{2}-1}}\left[\left(\boldsymbol{L}_0(2 |\beta|)+\frac{1}{\pi  |\beta|}\right) K_1(2 |\beta|)-\frac{1}{2|\beta|}-\frac{1}{2\pi |\beta|^{2}}\right. \\
    &\left.+\left(
   \boldsymbol{L}_{-3}(2 |\beta|)+\frac{2}{|\beta|}\boldsymbol{L}_{-2}(2 |\beta|)+\frac{1}{2 \pi |\beta|^{2}}\right) K_0(2 |\beta|)\right].
\end{aligned}
\end{equation}
This evaluated form is immediately amenable to the massless limit, giving
\begin{align}
    \mathcal{F}^{(2)}_{2a}(|\beta|)\overset{|\beta|\rightarrow0}{\longrightarrow}-\frac{1}{2\sqrt{\gamma^{2}-1}}\,,
\end{align}
in agreement with \cref{eq:FTMasslessLimits}.

\end{widetext}

\bibliography{ScalarWQFT-v2}

\end{document}